# Increasing Well-being through Robotic Hugs

Oliver Bendel [1], Andrea Puljic [1], Robin Heiz [1], Furkan Tömen [1], and Ivan De Paola [1]

[1] *School of Business FHNW, Bahnhofstrasse 6, 5210 Windisch, Switzerland*

**Abstract**

This paper addresses the question of how to increase the acceptability of a robot hug and whether such a hug contributes to well-being. It combines the lead author's own research with pioneering research by Alexis E. Block and Katherine J. Kuchenbecker. First, the basics of this area are laid out with particular attention to the work of the two scientists. The authors then present HUGGIE Project I, which largely consisted of an online survey with nearly 300 participants, followed by HUGGIE Project II, which involved building a hugging robot and testing it on 136 people. At the end, the results are linked to current research by Block and Kuchenbecker, who have equipped their hugging robot with artificial intelligence to better respond to the needs of subjects.

## 1. Introduction

Hugs have been part of human interaction since time immemorial. Descriptions can be found, for example, in Ovid's "Metamorphoses" [1]. The mountain nymph, Echo, approached her beloved Narcissus with outstretched arms. He, however, withdrew from her embrace. The unfortunate woman then hid in a cave and spurned food until she consisted only of voice. We hear her when we raise our voices even in a cave or in the mountains [2]. The Cypriot sculptor, Pygmalion, who had fallen in love with Aphrodite but could not reach her as a mortal, created a sculpture resembling her, which he placed in his bed and embraced. The goddess of love and beauty saw the couple. She took pity on him and brought his sculpture to life. It did not end at a hug: Pygmalion even had offspring with Galatea, as she was now called [3].

Hugs are important for the well-being of people of all ages [4]. However, not everyone can be hugged by someone or hug someone, for example, because they are lonely, single, or isolated. According to the American psychotherapist Virginia Satir, we need – as she is quoted in numerous media – four hugs per day for survival, eight hugs per day for sustenance, and twelve hugs per day for growth [5]. Even if such numbers should not be taken entirely seriously, they underscore the importance of human interaction for well-being (including satisfaction, calming, and happiness) [6, 7].

For some years now, people have been fascinated by social robots [3]. They also have a long history of ideas and, like R2-D2 or Marvin, appear in science-fiction films and books. Their actual proliferation began in the 1990s, partly through developments at the Massachusetts Institute of Technology (MIT) [3]. Since 2010, there has been a real boom, although the market is still not stable [8]. Social robots such as NAO, Pepper, Cozmo, Paro, and Lovot, or Harmony and Sophia, are now widely known.

In the course of dealing with humanoid robots, Alexis E. Block and Katherine J. Kuchenbecker (ETH Zurich and Max Planck Institute for Intelligent Systems) began to question the acceptability of a robot hug. Starting in 2018, they used PR2, which they introduced to subjects as HuggieBot and which underwent several stages of development over time [9, 10, 11]. The lead author, inspired by this research, initiated the HUGGIE projects I and II at the School of Business FHNW [12, 13]. A dedicated hugging robot was to be built as early as 2020, but this was thwarted by the restrictions imposed in the wake of COVID-19. Instead, a large online survey was conducted (HUGGIE project I). In 2022, HUGGIE could then be built and tested (HUGGIE project II). This robot would be a human-sized, human-like figure with a synthetic voice.

Of course, human hugging is in no way to be abolished or relativized. Rather, the authors' consideration is that extreme situations can occur where human touch and hugs are not present or not present in sufficient quantity resulting in well-being deficits. Examples include wars, disasters, pandemics, and lunar and Martian flights. However, robotic hugging could also be an option in environments where there is a lot of hustle and bustle and stress, such as shopping malls or open-plan offices [14]. Last but not least, following on from the above, lonely and single people could be emotionally cared for in this way, at least temporarily or additionally.

Obviously, robotic hugs are not the only way to increase people's sense of well-being. In nursing homes and homes for the elderly, social robots are used that can be touched and stroked (a famous example in the tradition of animal therapy is Paro) or with which people can talk. In addition, there are stuffed animals and pillows of all kinds that are perceived as pleasant. Some of them are soft and warm or play music. For the sake of brevity, these options cannot be discussed.

The present paper is devoted to the question of how the acceptance of a hug by a social robot can be increased and whether such a hug can contribute to well-being. It combines the lead author's own research with the pioneering studies of Block and Kuchenbecker. First, it lays down the groundwork for this area, focusing in particular on the work of the two scientists mentioned. Then the authors present HUGGIE Project I, which essentially consisted of a survey with nearly 300 participants, followed by HUGGIE Project II, in which a hugging robot was built and tested on 136 people. At the end, the results are linked to current research by Block and Kuchenbecker, who have equipped their hugging robot with artificial intelligence to better respond to the needs of the subjects [11].

## 2. Research on Hugging Robots

Social robots are robots created to interact with humans or animals. They interact and communicate with living beings and are close to them [3]. They are new entities in the social fabric. Many of them are designed to be humanoid or animaloid, thus replicating aspects of living beings. The utility does not necessarily lie in the useful and social, but in practice it often does when thinking about care and therapy. Hugging robots are mostly humanoid in design and equipped with arms.

Research on robotic hugging certainly exists, for example in the context of models such as The Hug, ARMAR-IIIb, Robovie, Telenoid, and Hugvie [15, 16, 17, 3]. However, some studies do not focus on being hugged by a robot, but on hugging a robot (or a figure), which is a significant difference. Figures like Telenoid and Hugvie from Hiroshi Ishiguro Laboratories have arms that are far too short to wrap around the body of their human counterparts, not to mention that they cannot move their arms of their own accord. But they can certainly be hugged, unlike small cuddly toys. When one is hugged, one usually returns the hug, so it can be said that both possibilities occur here at the same time. In rare cases, one rejects the embrace. Block and Kuchenbecker had both hugging possibilities in mind, as did the lead author at the School of Business FHNW, who initiated and supervised the HUGGIE projects.

Based on previous research on the topic of robot social interaction, Block and Kuchenbecker discovered that humans find interaction with a robot more enjoyable when it initiates the conversation and shows engagement, for example, through gestures [9]. Furthermore, it seems to be of particular importance that a robot mirrors the behavior of its human counterpart, known as "The Chameleon Effect" [2]. Therefore, the researchers designed their experiment according to these findings and tested whether a humanoid robot that is touch-sensitive, warm, and soft can hug humans in a satisfactory manner by modifying hug pressure and duration.

They investigated the importance of warmth and softness during a hug by covering the PR2 robot with foam and fabric and attaching heating elements to its exterior [9]. In addition, the height and contact pressure of the robot were adjusted to the physical characteristics of each subject. To ensure an appropriate duration of the hug, haptic sensors were used. This allowed the start and end of the hug to be determined by measuring the intensity of human contact [2].

Block and Kuchenbecker tested robot hugging under three experimental conditions: hard-cold, soft-cold, and soft-warm [2]. More specifically, robot PR2 was either in its original state with no additional material (hard-cold), wore layers of fabric and foam (soft-cold), or had heated cushions sewn into the fabric (soft-warm). 30 subjects hugged PR2 aka HuggieBot in the soft-warm condition during the experiment [9].

The experiment showed that the foam and fabric layer increased the subjects' comfort and perceived safety during the hug, leading to the conclusion that people preferred to be hugged by a soft robot rather than a hard one. However, the most significant effect on perceived safety and comfort came from the addition of heat. Thus, the two researchers concluded that a warm robot is preferred over a cold one in the context of hugs. Softness, warmth, and responsiveness are factors that significantly affect the quality of robot hugs [9].

Insights were also gained into the pressure and duration of the hug, whereby average values of human hugs have been known for some time. Initially, the problem remained that every human reacts differently and it can happen with individuals that they find the robot unpleasant despite the selected average hug duration. An adaptive solution was needed here, which led to the addition of artificial intelligence (AI), especially machine learning [11].

## 3. HUGGIE Project I

The HUGGIE Project was started at the beginning of 2020, and was subsequently renamed HUGGIE Project I to distinguish it from its successor project [12]. Originally, the student team was supposed to build and test its own hugging robot. However, COVID-19 made testing impossible, as the test subjects could have become infected. In general, operations at the university were largely suspended for months, and for a time it was only possible to enter the buildings with a special permit.

The lead author, as the client and supervisor, agreed with the two students that instead of developing a robot, a broad online survey should be conducted. This was to produce findings on attitudes towards robot hugs and to show what factors influence acceptance. The aim was to look for differences between men and women and to find out how people would behave towards robots with gender characteristics.

The online survey targeted German-speaking European residents, e.g., from Switzerland, Germany, and Austria [12]. A total of 337 participants opened the survey advertised via social media, of which 287 completed the survey. It is important to note that the sample drawn is not representative of the entire population in German-speaking countries, neither in terms of sample size nor in terms of diversification of participants.

Most participants had a positive attitude towards robots [2]. For some of them, however, a physical interaction with a social robot meant crossing a certain boundary. Social robots are apparently perceived positively when observed from a sufficient distance, but closer contact or physical interaction with them is more hesitantly accepted. The main reason for skepticism about hugs by a robot is that it is perceived as a device rather than an equal partner.

The data analysis showed that women prefer a same-sex hugging partner, while men prefer to be hugged by a person of the opposite sex. This suggests that both men and women prefer to be hugged by a woman. Interestingly, the same tendency was observed with respect to being hugged by a robot if there was a gender preference. In this case, both men and women prefer to be hugged by a robot with female attributes [2]. It should be added that most participants did not have a gender preference for robots and could imagine approximately gender-neutral forms.

The majority of participants wanted the robots to have a human-like appearance, albeit a cartoonish appearance like Pepper rather than a highly realistic one like Sophia, which was largely rejected in this context. Other features that could contribute to a hugging robot's appeal include "soft" vibrations that mimic a heartbeat (similar to Hugvie) and a pleasant scent (one participant spoke of wanting the robot to smell like chocolate). Eighty percent said a social robot's ability to talk and make sounds was important.

In addition, the robot should be height adjustable. According to the survey, the majority of respondents prefer to hug someone who is more or less at eye level with themselves, and this of course depends on the individual user. Among other things, a gender analysis of the data revealed that female participants preferred to hug someone taller than themselves. Male participants preferred a counterpart who was smaller than them. This is likely to be transferable to robots.

Regarding the characteristics of the robot's arms, the survey results suggest that warm and soft arms are preferred over hard and cold arms when hugging. Block and Kuchenbecker's experiment produced the same result [9]. Interestingly, male survey participants indicated a stronger preference for warm arms than females.

When it comes to the "inner values" of a social robot, the ability to show empathy seems to play an important role. Various experiments have confirmed that humans treat robots with a similar degree of empathy as fellow humans [18] and some probably expect robots to do so as well. What is surprising here is that the display of emotion seems to be less important than the ability to be empathic.

HUGGIE Project I was completed in August 2020. The basic principles and results were recorded in a book chapter published under the title "In den Armen der Maschine" ("In the arms of the machine") in the book "Soziale Roboter" ("Social Robots") [2]. Within a short time, the book became a standard work on social robotics in the German-speaking world.

In addition to HUGGIE Project I, two other projects were conducted that are relevant in this context. In one, robots were sketched that could be created from soft, simple shapes [19]. In this project, Hugvie was included and modified on paper – a male version was adapted from the rather female-looking classic version. Another project on robot enhancement systematically discussed how to extend and improve an existing social robot, such as adding wigs, clothing, accessories, silicone masks, and programming enhancements [20]. This is exactly what is taking place in the projects around HuggieBot.

## 4. HUGGIE Project II

In the summer of 2022, the School of Business FHNW invited tenders for another HUGGIE Project, which was named HUGGIE Project II to distinguish it from its predecessor. A team of four was recruited, with Andrea Puljic, Robin Heiz, Furkan Tömen, and Ivan De Paola. This time it was not a thesis but a practical project with even more manpower available [13]. The task was again to build and test a hugging robot. This should be based on the results of HUGGIE Project I. The aim here was to test whether a robotic hug contributes to well-being, and whether voice and vibration as a simulation of the heartbeat increase the acceptance of a robotic hug. Thus, suggestions from the survey in HUGGIE Project I were implemented.

### 4.1. Project Preparation

In the first meeting, the requirements were specified by the client [13]. He required two prototypes. The main one was HUGGIE, the secondary one, which was to serve as a comparison, would be a conventional teddy bear (called Teddy here). A budget of 500 CHF was available. The minimum requirements regarding HUGGIE were as follows:

- HUGGIE should have a human-like appearance.
- It should have movable arms with which it can hug someone.
- It should be padded and thus soft.
- It should radiate warmth in certain places.
- It should have a vibrating element that mimics a heartbeat.
- It should have an integrated voice assistant with a synthetic voice.

Additional functions could be added during the project, according to the agreement, and when agreed with the client. The requirements for Teddy were:

- Teddy should allow a comparison with HUGGIE.
- It should be a life-size stuffed teddy bear.
- It should not have the characteristics of a social robot.

Additional functions could be added during the project, again in consultation with the client.

### 4.2. Development of HUGGIE

First, there was a brainstorming session between the client and the team on how HUGGIE should be built [13]. With a budget of 500 CHF, a real robot of the required size could not be obtained. It was decided to simulate a social robot, but to use technical components found in real robots, such as synthetic voice or vibration.

A height-adjustable dressmaker's dummy was procured via online mail order, with a human-like head, upper body, arms, and a metal frame. The figure was equipped with a woolen cap and a gray hoodie. Invisible strings were attached to the arms, which could be used to move them. The front area was padded, as were the arms. The upper body was fitted with an electric blanket. So that the electric blanket held, it was fastened with a belt.

The overall aim was to create a figure that reflected the findings of Block and Kuchenbecker, had a humanoid appearance, and appeared as neutral as possible in terms of gender and age. The interchangeable or switchable elements included the vibration in the heart area and the voice assistant. During the course of the project, it was agreed that the request of the participant in the HUGGIE Project I survey should also be accommodated and a scent of coconut (instead of the smell of chocolate) applied. These elements are discussed in more detail below.

Speakers were placed in the head area of the dressmaker's dummy to transmit the voice of the voice assistant. The voice assistant was the Speechify application from the Apple AppStore, with the voice Stephanie UK selected. The pleasant sounding voice could be played during the hug, saying, "Oh, hey! Good to see you. Can I give you a hug?" … These phrases had been agreed upon during a meeting between the client and the team.

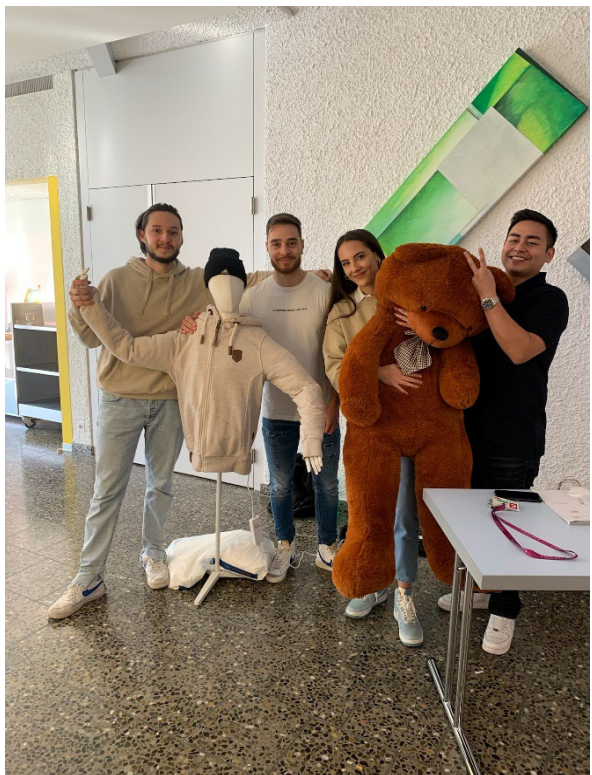

**Figure 1**: Team with HUGGIE and Teddy

The scent was Coconut Body Mist from The Body Shop brand. The coconut note was unmistakable and sort of bridged the gap to a food product. The team members sprayed the substance onto a regular cotton pad and wrapped it in a plastic bag. They pierced it several times with a needle to allow the scent to spread. The application was then attached to HUGGIE's cap with a safety pin where needed.

The smartphone with vibration was placed behind the electric blanket, in the middle of the chest, supported by the belt. Thus, it could not fall off, and the hardness of the smartphone was concealed by two layers (hoodie and electric blanket). An app from the Apple AppStore was also used, called iVibrate – Vibrator Phone Spa. There, in the free version, one could choose different massage patterns. The team took Vibes with level 3/6, which was two strong vibrations followed by a small vibration, imitating a heartbeat pattern.

### 4.3. Testing of HUGGIE

Before the official testing began, a pre-test was conducted in a Swiss energy and automation technology company (known for its industrial robots, among other things) [13]. There, a total of 64 participants were hugged by HUGGIE and gave their feedback on it. The number is very high for a pre-test and came about simply because the employer of a team member made it possible to access a larger group at short notice.

The result of the pre-test showed which properties HUGGIE absolutely had to have in order for the hug to have a pleasant effect. The warmth proved to be indispensable here – as Block and Kuchenbecker's research had already suggested. The lack of padding on the back was repeatedly objected to by the participants and was therefore corrected with the help of additional padding. In consultation with the client, the survey items were also adjusted after the pre-test and height, gender, and age were added. This was done to obtain information on whether or not HUGGIE was better received by an age group or gender. In addition, it was important to know the height of the respondent for the height of HUGGIE to be perceived as comfortable or uncomfortable.

The final test with 72 subjects took place at another Swiss company in the energy sector (42) and at the School of Business FHNW (30). There were about the same number of men as women. One person indicated "non-binary" for gender. Most were between 21 and 45 years old. Only one was over 65, which can be considered a problem.

The test was performed in four different sequences with four different hugs. This was to avoid distorted results from the possible skepticism of the unknown or increasing habituation. The four versions differed as follows:

1. Base
2. Base + voice
3. Base + vibration
4. Base + voice + vibration + scent

Further combinations were refrained from, even if they would have made sense. The background was that during the pre-test several participants had indicated that two hugs were already too much for them. This was – in addition to the research results already available – another reason for adding warmth to the basic version alongside softness and offering three variations based on this.

Each participant was hugged four times by HUGGIE. After that, another hug was performed for comparison with the teddy, whose arms were moved from behind by a team member. Most of the subjects returned the hug, thus wrapping their arms around HUGGIE. Some were skeptical about being hugged by it, and some even refused at first. One opinion expressed several times was that a robot can never replace a human being, since the latter shows emotions during a hug, which are necessary for a pleasant sensation. Immediately after the hugs, each participant was subjected to a survey, using a structured online questionnaire. In addition, observations took place, but without prior determinations.

The survey was divided into 11 parts. In sections 1 to 4, information was obtained about the participant. In section 5 the question was "Did you feel any differences between the four hugs?", in section 6 "If yes, which hug was the most comfortable?". Section 7 was devoted to well-being during the hug. The aspects of calming, happiness, and satisfaction in the broadest sense were of interest. In addition, questions were asked about whether the hug was perceived as weird. In section 8, questions were asked about the robot, including warmth, softness, height, and voice. Section 9 compared HUGGIE and Teddy, and section 10 examined HUGGIE's presence in the workplace or at home. In section 11 ("Any other concerns you would like to share?") free text entry was possible.

Between 50 and 70 percent of pre-test participants agreed with the following statements: "The hug calmed me down" (requirement for calming), "The hug made me happier" (requirement for happiness), and "The hug helped me in some way" (requirement for satisfaction). Most of them checked "Agree", and a few checked "Agree completely". At the same time, over 50 percent agreed with the statement "The hug felt weird", suggesting that a robot hug is perceived as unusual, even strange [21]. This is in line with the results of HUGGIE Project I. Some may also have rejected a robot hug outright.

In the main test, the situation was initially similar. About 60 percent agreed with the statements "The hug calmed me down" and "The hug made me happier". Most of them ticked "Agree", a few "Agree completely". There were significant losses for "The hug helped me in some way", which only about 40 percent still underlined. The number of those who supported "The hug felt weird" even rose to just under 70 percent. This can perhaps be explained by the fact that there was less affinity for robots in the sectors (energy and education).

In the following, only the main test will be discussed [13]. A recurring comment was that HUGGIE should be softer (despite it already being made softer compared to the pre-test) but not as soft as the teddy bear. In addition, it was found that the height plays an important role in the hug; at best, it should be able to adapt to the client. The subjects were more satisfied when HUGGIE was the same height or slightly taller than they were. The fixed height of 175cm was acceptable to the majority, as most subjects were between 171 and 180cm tall. This largely confirms the results of HUGGIE Project I. In the survey of HUGGIE Project II, HUGGIE's warmth also proved essential. Except for a few outliers, the vast majority voted that it should not be colder than 47°C (due to the different layers the heat on the surface is reduced). This confirms the results of Block and Kuchenbecker and of HUGGIE Project I.

It turned out that just under half of the participants found the voice pleasant, and a good half found it unpleasant. About 75 percent denied that a male voice would be better. This suggests that a female voice during the hug is most likely to increase acceptance of a robotic hug. Female voices are generally preferred by manufacturers in voice assistants and female robots seem more likely to be trusted than male ones [3]. The content of what is said may also play a role. Care was taken to ensure that sentences seemed coherent and did not distract participants too much from the hug.

Opinions were equally divided on the vibration, which was intended to simulate the heartbeat. Through feedback, it was determined that participants do not feel the heartbeat in a normal hug and therefore the vibration can be somewhat irritating. This is an important point and suggests that one does not necessarily need to use a heartbeat, and if one does, it should be in a discrete and natural form. This incidentally raises questions in relation to Hugvie, where there is also a vibration element in the original Japanese version (this is intended to simulate the heartbeat of the human communication partner).

Opinions were just as different about the coconut scent that had been added. According to feedback, a classic perfume would be better but this should be personalized, as it allows one to cater to personal preferences. However, this would be logistically difficult to implement in practice. One possible way out, which admittedly involves many problems, is to try electrical stimulation of the brain to create the impression of smells [22].

The average length of the hug lasted three to five seconds. This proved to be pleasant for the most of those involved in the survey. However, observations showed that as soon as the customer himself or herself eased the pressure, HUGGIE should release the hug so that the feeling of well-being was maintained. Accordingly, the customer should basically be able to decide for himself or herself how long the hug should last and what his or her own preference is.

It was also asked which hug felt better, that of HUGGIE or that of the teddy bear. The results showed that there was no clear favorite. However, a slight majority voted for HUGGIE. For many, the teddy bear is a childhood memory and therefore seems familiar. However, people are not used to being hugged by a teddy bear, which may have weakened the positive effect somewhat.

Further it was asked whether one could imagine having HUGGIE at home or at the workplace. This was vehemently rejected. Especially when asked if they would like to have their own HUGGIE at home, there were only a few yeses. When asked why this was unthinkable, it came out that many subjects found the human hug indispensable [6]. Some subjects even associated a specimen of HUGGIE in their own homes with psychological problems.

In summary, it can be said that a considerable number of subjects benefited from being hugged by HUGGIE. They stated in the questionnaire that the hug triggered a sense of well-being in them; observations and conversations were able to confirm this. The addition of softness and warmth to HUGGIE's basic version contributed decisively to this. Less important were voice, vibration, and smell – all versions were equally favored. However, a female voice was strongly preferred. Some subjects seemed to have a downright reluctance to be hugged by a robot. This is consistent with the results of HUGGIE Project I and indicates that touch by robots should not be forced under any circumstances and should be an exception.

## 5. HuggieBot with AI Extension

HUGGIE could benefit from artificial intelligence when referring to Natural Language Processing (NLP). It is obvious to include other possibilities, for example, to analyze the subject's state of mind and adapt the system accordingly. Artificial intelligence for exactly this purpose was in the foreground in the latest phase of the HuggieBot project with HuggieBot version 3.0 [11].

HuggieBot and HUGGIE are similar in some aspects. Both are padded and have thermal elements – the team at the School of Business FHNW took their cue from the findings of the Max Planck Institute and ETHZ in this regard. There are coincidental similarities, such as the gray hoodie. Also, the realization that a synthetic voice could be useful probably developed independently. The lead author has been working on this topic for some years [23].

At the same time, there are significant differences. For example, HuggieBot in version 3 has a box-like head with a display showing eyes and mouth. Eyes, mouth, and other features mimicking a human appearance, on the other hand, were deliberately omitted in HUGGIE. The face was not one of the factors to be tested, and no unnecessary affection or rejection was to be generated, meaning that painted or glued-on eyes – or any other facial features – were not used.

An article in MIT Technology Review outlines the use of the AI components: "The robot tested is already version 3.0 – sensors measure pressure and noise in the robot's rear, air-filled torso. With this data, the team has trained a machine learning system to recognize people's desires." [11] The journalist who tested the robot reported: "And that works. I end the embrace slowly – as if I wanted to break free from the embrace of a human being – and notice: The robot obviously senses that I have had enough now and opens its arms as well." [11] This is an important improvement that was also addressed in HUGGIE Project II, where it was solved to some extent by the team's manual intervention.

Block and Kuchenbecker went far beyond survey and observation: "The researchers also measured people's physical responses, such as heart rate variability and levels of oxytocin and cortisol in saliva. 'The results are still being published', says MIP-IS Director Katherine Kuchenbecker. 'But we clearly see that the robot hugs lower stress levels.'" [11] For such an experimental setup, one needs the appropriate equipment, which is not available at the School of Business FHNW. However, it is conceivable that progress can be made in this regard through cooperation.

## 6. Discussion

Robotic hugs could be an option in the event of crises, catastrophes, and pandemics, as well as for flights to the moon, Mars, or other manned space flights. Likewise, lonely and single people could be helped temporarily. However, as the two HUGGIE Projects showed, not all subjects seem to agree with such a hug. In the current experiment, this may be because not all knew what to expect. Better explanatory work could be done here. However, in the first HUGGIE project, there was also strong evidence of basic refusal. In HuggieBOT, these problems seem to have occurred less [11]. Perhaps better explanations were provided or perhaps individual or cultural differences exist, such as differences in age. In principle, it is important to take the concerns of the subjects seriously.

The concept of robot enhancement was not applied in the actual sense in HUGGIE, since an existing social robot is assumed [24]. However, there is definitely a relationship, since different additions were made during the various runs. In the case of HuggieBot, one can easily speak of robot enhancement, namely with regard to the AI enhancement. Other additions and supplements would be worth testing.

Artificial intelligence could be used not only to analyze the behavior of the test subjects and to recognize their wishes and, based on this, to adapt the behavior of the robot, but also, for example, to give the robot a personality and let it show empathy and emotions in an advanced form. This raises new challenges, such as fraud and deception concerns [8].

With regard to the voice assistant, one can also go further and test other voices, ways of speaking, and content. For example, it would have to be clarified in more detail whether different female voices have different effects and how what is said can irritate rather than calm someone. In addition, it is possible to use fully functional voice assistants, e.g., based on GPT-3, which can conduct conversations before, during, and after the hug and thus strengthen the bond.

Research on hugging robots could also advance the field of care and therapy robots [3] as well as love dolls and sex robots [25]. In the case of care and therapy robots, touching and hugging have hardly been used so far. In sex robots, there are mainly hugs in the opposite direction, since moving arms are mostly missing. As soon as such would be available, hugs by the robot would be a logical consequence. Besides the approaches of hugging discussed, there is another that might be of interest that has been implemented with the Sense-Roid [26]. Here, the hug of a figure is personalized to the person hugging it with the help of sensors and actuators (vibrators and artificial muscles). In addition, it would be interesting to include projects on sleeping pillows and music pillows. In the case of music pillows, there are those where music is generated by AI, such as inmu (inmutouch.com), that could be integrated into hugging robots.

## 7. Summary and Outlook

This paper addressed the question of how to increase the acceptability of a robot hug and whether such a hug contributes to well-being. It combined the lead author's own research with the research of Alexis E. Block and Katherine J. Kuchenbecker and presented individual phases of the different projects.

First, the basics of this area were laid out, with particular reference to the work of Block and Kuchenbecker. In addition, Hugvie from the Hiroshi Ishiguro Laboratories and two of the lead author's projects were mentioned that have crossover with the HUGGIE projects. The authors then presented the HUGGIE Project I, which consisted mainly of a survey because of the pandemic, followed by the HUGGIE Project II, in which a hugging robot was built and tested.

In principle, it can be said that people benefit from robotic hugs and that they can increase their well-being as a result. In the pre-test there was clear evidence for this and sufficient evidence in the main test. However, as HUGGIE Project I revealed, some already have an aversion to robotic hugs in their imagination, which in turn some also showed in reality (HUGGIE Project II). Moreover, in the main test, the majority felt that the hug did not help them that could be interpreted as a contradiction – or shows that the question was inaccurately formulated. Such different needs must be taken into account.

It has already been highlighted that older individuals were largely absent from the subject group. Perhaps if older or lonely subjects were targeted for hugging, acceptance would be much higher. This possibility did not exist in this project but it would be interesting to start a follow-up project with residents of nursing and retirement homes. Drug addicts and other groups with social problems could also be included.

In line with research of Block and Kuchenbecker, warmth and softness of body and arms are vital. Voice, vibration, and scent were found to be less relevant. However, it would be interesting to simulate a weak, realistic heartbeat and to vary in scents and smells (both perfumes and food smells) and repeat the tests. In addition, it was clearly stated that a male voice would not be better, which provides an opportunity to focus on optimizing the female voice used to have a significant positive impact on the acceptance of the hug.

In research at the School of Business FHNW, it becomes clear that hugs are something very individual. Letting a robot hug a test person is initially comparable to letting a stranger hug a test person. The hug is not necessarily perceived as fitting. And whether one finds it pleasant, even after some habituation, depends on many different, individual factors.

In turn, this paper argues in favor of personalizing the robot. This is exactly what Block and Kuchenbecker are doing in their research. They combine general findings with individual ideas and expectations. In this way, hugging robots will emerge that have a high level of acceptance. However, they will also incur high costs, both in acquisition and in operation. Whether these pay off or can be borne depends in turn on the area of application.